\documentclass[sigconf]{acmart}
\AtBeginDocument{%
  }

\setcopyright{none}
\renewcommand\footnotetextcopyrightpermission[1]{} 
\begin{document}
\title{Towards AI Agents Supported Research Problem Formulation}

\author{Anrafel Fernandes Pereira}
\affiliation{%
  \institution{PUC-Rio and Univassouras}
  \city{Rio de Janeiro}
  \country{Brazil}}
\email{afpereira@inf.puc-rio.br}

\author{Maria Teresa Baldassarre}
\affiliation{%
  \institution{University of Bari}
  \city{Bari}
  \country{Italy}}
\email{mariateresa.baldassarre@uniba.it}

\author{Daniel Mendez}
\affiliation{%
  \institution{Blekinge Institute of Technology and fortiss}
  \city{Karlskrona}
  \country{Sweden}}
\email{daniel.mendez@bth.se}

\author{Marcos Kalinowski}
\affiliation{%
  \institution{PUC-Rio}
  \city{Rio de Janeiro}
  \country{Brazil}}
\email{kalinowski@inf.puc-rio.br}

\renewcommand{\shortauthors}{Pereira \textit{et al.}}

\begin{abstract}
[Background] Poorly formulated research problems can compromise the practical relevance of Software Engineering (SE) studies by not reflecting the complexities of industrial practice. [Aims] This vision paper explores the use of artificial intelligence (AI) agents to support SE researchers during the early stage of a research project: the formulation of the research problem. [Method] Based on the Lean Research Inception (LRI) framework and using a published study on code maintainability in machine learning as a reference, we developed a descriptive evaluation of a scenario illustrating how AI agents, integrated into LRI, can support SE researchers by pre-filling problem attributes, aligning stakeholder perspectives, refining research questions, simulating multiperspective assessments, and supporting decision-making. [Results] The descriptive evaluation of the scenario suggests that AI agent support can enrich collaborative discussions and enhance critical reflection on the value, feasibility, and applicability of the research problem. [Conclusion] Although the vision of integrating AI agents into LRI was perceived as promising to support the context‑aware and practice‑oriented formulation of research problems, empirical validation is needed to confirm and refine the integration of AI agents into problem formulation.
\end{abstract}



\maketitle
\section{Introduction}
A well-formulated research problem is fundamental to achieving practical relevance in Software Engineering (SE) research. However, studies show a persistent gap between the scope of academic contributions and industrial needs, often attributed to poorly formulated research problems and the lack of structured approaches to evaluate them~\cite{garousi2020, winters2024}. Technology transfer models, such as the one proposed by Gorschek \textit{et al.}~\cite{gorschek2006, gorschek2021solving}, provide valuable guidelines for co-production between academia and industry, but do not cover the early stages of problem formulation. Moreover, most existing efforts focus on evaluating research results~\cite{ivarsson2011, molleri2023, petersen2024}, leaving an open need for structured methods to formulate relevant problems from the outset.

Pereira \textit{et al.}~\cite{PaperEASE2025} advanced in this direction by proposing a framework called Lean Research Inception (LRI), designed to support the early stages of SE research, guiding the formulation and initial assessment of research problems from the perspective of industrial relevance. The LRI was developed to enable a structured collaboration between researchers and practitioners, with the goal of ensuring that research problems are grounded and reflect the complexities of industrial practice~\cite{marijan2020}. Its foundations are based on agile methodologies, including Design Thinking~\cite{plattner2009}, Lean Startup~\cite{ries2011}, and Lean Inception~\cite{caroli2017}, which emphasize iterative development, early and continuous stakeholder involvement, and the delivery of practical value to industry. The LRI framework is publicly available\footnote{Miro Template for Lean Research Inception: \url{https://miro.com/miroverse/lean-research-inception-template/}}, encouraging adoption by the SE research community.

In this study, we explore the vision of integrating Artificial Intelligence (AI) agents to support SE researchers in the research problem formulation phase. Taking the LRI framework~\cite{PaperEASE2025} as a basis, we explore the feasibility of integrating these agents into the methodology. We argue that such integration constitutes a promising opportunity to improve the formulation of research problems of practical relevance and to broaden the holistic understanding of the complex phenomena in the area. To this end, we discuss how AI tools, such as Google’s AI Co‑Scientist~\cite{gottweis2025}, can reinforce early research activities by capturing relevant information, facilitating collaborative discussions among professionals from different backgrounds, and providing preliminary multiperspective assessments. Based on a descriptive evaluation of a practical scenario involving Machine Learning (ML) projects~\cite{cabral2024}, we show how AI agents integrated into the LRI framework can support the collaborative formulation of research problems. With AI support throughout the entire LRI stages, we envisage a more interactive and practice‑oriented approach to SE research.

\section{Background and Related Work}
The challenge of achieving (and demonstrating) practical relevance in Software Engineering (SE) research and the lack thereof in many contributions has been widely discussed \cite{garousi2016, winters2024}. Criticisms include a disconnect from industrial practice, narrowly scoped studies, and simplistic views of real-world challenges. Bridging this gap requires strategies that not only align research efforts with industrial realities, but also balance theory with real-world applicability \cite{stol2018}.

Ivarsson \textit{et al.}~\cite{ivarsson2011} proposed a model to assess the rigor and industrial relevance of technology evaluations. Applying it to a systematic review of requirements engineering studies, they found that most of them lacked both rigor and relevance, with no observable improvement over time. Garousi \textit{et al.}\cite{garousi2020} conducted a multivocal literature review (MLR) to assess the practical relevance of SE research. They identified three main limitations: simplistic views of practice, weak ties with industry, and poor problem formulation. To address these issues, they recommend using context-aware research methods, focusing on real-world problems, and strengthening collaboration with industry. Winters\cite{winters2024} reinforces this view by highlighting the lack of practical context, narrow problem scopes, and limited scalability in SE research, advocating for results that deliver measurable value to industry.

Petersen \textit{et al.}~\cite{petersen2024} propose a reasoning framework to improve the design, assessment and reporting of industrial relevance in SE research. Given the lack of consensus on the definition and measurement of relevance, they review key attributes such as applicability, context, and practical impact. Their framework, structured around six aspects \textit{(what, how, where, who, when, and why)}, provides an approach to evaluate and communicate the relevance of applied research research in industry contexts. Storey \textit{et al.}~\cite{storey2024disruptive} propose a practical and provocative methodological approach to support researchers in formulating questions about the impact of disruptive innovations in SE. The model emphasizes interdisciplinary collaboration, industry engagement, and iterative experimentation, encouraging a sociotechnical reflection on research objects, theories, and methods.

Marijan and Gotlieb~\cite{marijan2020} highlight the benefits of fostering collaboration between researchers and practitioners through the co-creation model, emphasizing that this process can foster the development of solutions more aligned with real-world needs. The authors also point out that co-creation can play a crucial role in formulating research problems, ensuring that research is guided by challenges relevant to professional practice. The authors focus on sharing lessons learned by offering practical reflections, but without providing an operational model to guide their application.

Pereira \textit{et al.}~\cite{PaperEASE2025} introduce Lean Research Inception (LRI), a structured framework aimed at aligning research with industrial relevance by means of a process that fosters collaboration between researchers and practitioners, focusing on the early research stage — problem formulation. LRI uses a visual board called Problem Vision, which organizes seven attributes. The attributes reflect the importance of defining the initial pain point or opportunity (i.e., the problem)~\cite{gorschek2006, shaw2002}, contextualization~\cite{runeson2012case, stol2018,  wieringa2014}, practical impact~\cite{lo2015}, practitioners’ identification~\cite{dyba2005}, empirical evidence~\cite{wohlin2013}, well-defined objectives~\cite{basili1988, wieringa2014}, and precise research questions~\cite{basili1988, shaw2002}. 

In addition, to support the initial assessment of the formulated research problem, LRI employs a semantic differential scale to evaluate value, feasibility, and applicability. These tools are articulated in five sequential phases — (1) \textit{Problem Vision Outline}, (2) \textit{Problem Vision Alignment}, (3) \textit{Research Problem Formulation}, (4) \textit{Research Problem Assessment} and (5) \textit{Go/Pivot/Abort Decision}.

These studies highlight the urgency of rethinking how research problems are formulated in Software Engineering~\cite{garousi2020, ivarsson2011}. They draw attention to persistent gaps between academic investigation and industrial practice, emphasize the need for systematic structures to assess practical relevance, and advocate for a sociotechnical collaborative and interdisciplinary reflection~\cite{petersen2024,storey2024disruptive}. In this context, Lean Research Inception (LRI) emerges as a structured framework that organizes problem formulation into seven attributes and applies a semantic differential scale for initial assessment, demonstrating potential to bridge research and practice~\cite{PaperEASE2025}.

This work explores the vision of integrating AI agents, such as Google’s AI Co‑Scientist~\cite{gottweis2025}, to support SE researchers into the stage of formulating research problems. This integration can complement frameworks like LRI by synthesizing knowledge, capturing relevant contexts, and offering multiperspective assessments in the initial stages. Recent evidence reinforces the potential for using AI to support researchers, for example, in systematic literature reviews ~\cite{felizardo2024chatgpt, huotala2025sesr}. By integrating AI agents into LRI’s collaborative process, we envisage a more agile, contextualized, and practice-oriented approach to research problem formulation, aligned with the goal of making SE research more relevant to industry.

\section{Exploring AI-Supported Research Problem Formulation}
Recent advances in Artificial Intelligence (AI), particularly in the development of Large Language Models (LLMs) and collaborative agents, have demonstrated significant potential to enhance scientific processes and methodological frameworks~\cite{felizardo2024chatgpt, gottweis2025, huotala2025sesr, storey2024disruptive}. A notable example is the AI Co-Scientist, developed by Google and introduced by Gottweis \textit{et al.}~\cite{gottweis2025}, which adopts a multi-agent asynchronous architecture to support iterative hypothesis generation, and progressive refinement of ideas, effectively simulating a collaborative scientific environment.

AI-based systems with these capabilities naturally align with methodologies that emphasize collaboration, such as LRI, which aims to foster the practical relevance of SE research from its early stages. Integrating agents like Google’s AI Co-Scientist~\cite{gottweis2025} into LRI can significantly enhance the quality of research problem formulation and assessment, enabling researchers and practitioners not only to access and synthesize technical knowledge more efficiently but also to critically reflect on the social, economic, and ethical implications of their investigations. With this view, formulation and initial assessment of research problems can be enriched by a reflective and context-aware approach, essential to address the contemporary challenges of applied research in SE.

\subsection{Integrating AI Agents into Lean Research Inception}
The vision of integrating AI agents in LRI aims to enhance the ability of researchers and practitioners to formulate and assess research problems with greater practical relevance. To explore the feasibility of this proposal, we developed an illustrative scenario as a descriptive evaluation~\cite{dresch2014design}. This scenario, inspired by a research problem presented by Cabral \textit{et al.}~\cite{cabral2024}, and also used in previous empirical studies~\cite{PaperEASE2025}, features a SE researcher leading a team at a technology company facing challenges related to code maintainability in Machine Learning (ML) projects.

\begin{figure}[h!]
    \centering
    \includegraphics[width=1\linewidth]{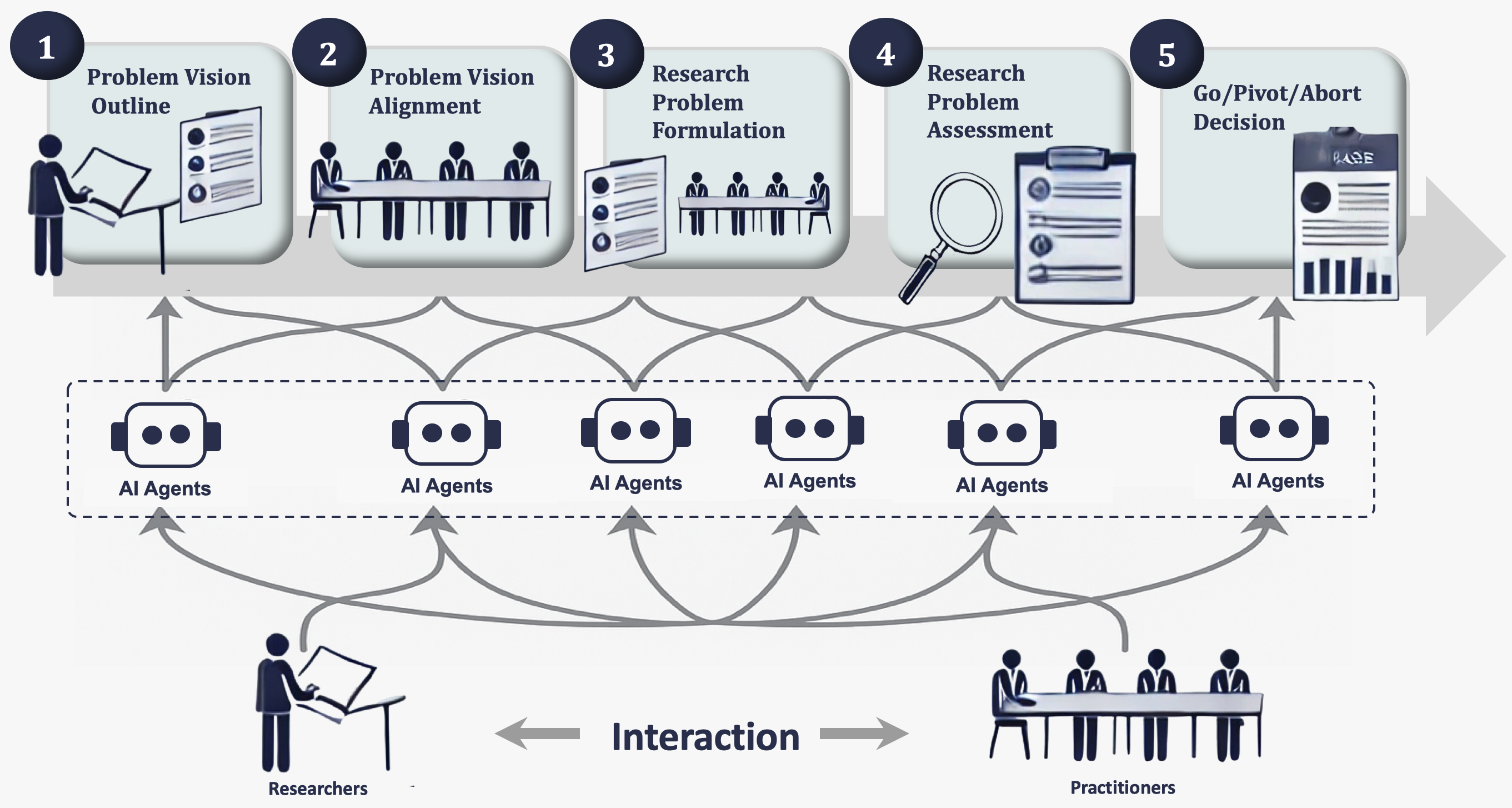}
    \caption{AI Agents Integrated with LRI}
    \label{fig:lri_ai}
\end{figure}

Based on this context, we described how AI agents could support each of the five phases of the LRI (presented in the first layer of the Figure~\ref{fig:lri_ai}), using practical examples to demonstrate how the integration can assist in pre-filling the seven attributes of the Problem Vision board, promote collaborative alignment across diverse professional profiles, and support the initial assessment of research problems through multiperspective feedback. Figure~\ref{fig:lri_ai} illustrates this vision by introducing an additional layer that shows how AI agents can be integrated into the different phases of the LRI, increasing the support provided throughout the process.

Although this descriptive evaluation cannot yet follow the structure of a formal experiment, it already provides a useful preliminary assessment, revealing promising opportunities for how AI can serve as intelligent support in the formulation of more contextualized and industry-aligned research problems.

\textbf{Phase 1 — Problem Vision Outline:} In the first phase of the LRI, the researcher prepares a draft of their research problem using the Problem Vision board. We envision the contributions of integrating AI agents at this phase as follows: the AI agent can explore large volumes of scientific literature, white papers, industry reports and domain‑specific repositories to identify information relevant to the researcher’s initial goal. These insights can help the researcher pre‑fill the seven attributes of the Problem Vision board (\textit{practical problem}, \textit{context}, \textit{implications/impacts}, \textit{practitioners}, \textit{evidence}, \textit{objective}, and \textit{research questions}) reducing cognitive load and accelerating the drafting stage.

\textit{Practical scenario:} After reporting that \textit{``data scientists produce ML code that is difficult to maintain,''} the researcher can receive a technical overview from the AI agent, including evidence from academic studies and industry cases describing similar issues. The agent can suggest pre-filled content for attributes such as \textit{practical problem}, \textit{context}, \textit{implications/impacts}, \textit{stakeholders}, and \textit{evidence}, including maintainability metrics, failure examples, and impacts observed in industrial projects.

\textbf{Phase 2 — Problem Vision Alignment:} In this phase, the researcher invites professionals with practical experience related to the problem under investigation to participate in a collaborative workshop. During this workshop, researchers and practitioners jointly review and discuss each of the seven attributes of Problem Vision to ensure a shared understanding and improve the formulation of the draft research problem. During alignment, the AI agent acts as a knowledge mediator, providing illustrative examples, related cases, and boundary conditions to enhance mutual understanding and guide collaborative problem co-formulation.

\textit{Practical scenario:} In a meeting with engineers and data scientists, the AI Co-Scientist can facilitate the discussion by presenting real code snippets and visualizations that highlight missing best practices. It can simulate how different professional profiles perceive the problem and helps all participants collaboratively discuss and refine the Problem Vision board, fostering consensus on the problem formulation.

\textbf{Phase 3 — Research Problem Formulation:} Based on alignment, the research problem is documented, serving as the basis for phase 4. At this stage, the AI multi-agent reasoning architecture provides support for iterative refinement of the problem. It can identify inconsistencies, suggest complementary research questions, and compare the formulated problem with similar challenges already addressed.

\textit{Practical scenario:} Based on previous alignments, the agent can suggest additional questions such as: \textit{``What design practices are commonly neglected by data scientists in industrial projects?''} or \textit{``How can code readability and scalability be improved in ML models deployed in production?''} It can identify gaps in the literature and recommends relevant methodological approaches.

\textbf{Phase 4 — Research Problem Assessment:} At this stage, each LRI participant is invited to assess the formulated research problem using a seven-point semantic differential scale, considering \textit{value}, \textit{feasibility}, and \textit{applicability}. The LRI´s semantic differential scale can be enriched through AI-generated stakeholder simulations. The agent can provide detailed justifications for assessments of \textit{value}, \textit{feasibility}, and \textit{applicability}, encouraging critical reflection even before engagement with industry partners.

\textit{Practical scenario:} The AI Co-Scientist can simulate the evaluations of a technical lead, a senior data scientist and a product manager. Highlights the strategic relevance of the problem, the availability of data for investigation, and the applicability in real-world contexts, providing rationale that reinforces the importance of the study.

\textbf{Phase 5 — Go/Pivot/Abort Decision:} After consolidating the assessments, it is decided whether the research continues, whether the problem should be adjusted (starting again in phase 2) or whether it should be finished, depending on the relevance and possibility of adjustment. In the decision-making phase, the AI agent can analyze risks, feasibility trends and scenario-based projections to suggest evidence-based strategic directions. Using previous case studies, failure patterns, and economic models, it supports decisions on whether to move forward, adjust, or abandon the proposed research.

\textit{Practical scenario:} Based on the evidence collected, the AI Co-Scientist can recommend continuing the investigation. The team acknowledges the clarity of the problem, its practical relevance, and potential industry impact, and decides to move forward with the development of a structured study on SE practices in ML projects.

At the end of the five phases, the SE researcher and their team not only arrive at a clearly structured and multiperspective assessed research problem, but also experience a new process mediated by AI. The AI Co-Scientist acts as a reasoning partner, not replacing researchers and practitioners, but enhancing their ability to perform critical analysis and decision-making. These agents can continuously interact with researchers, providing recommendations, explanations, and refinements, while also supporting practitioners through clearer evidence-based communication, strengthening the alignment between academia and industry.

In summary, the vision of integrating AI-based tools, such as Google's AI Co-Scientist~\cite{gottweis2025}, into the stage of formulating and initial assessment of research problems in SE was perceived as potentially helping to foster studies with greater practical relevance and to broaden the holistic understanding of the area’s complex phenomena. By harnessing AI agents’ ability to synthesize knowledge, anticipate stakeholders’ perspectives, and support critical reflection, the LRI framework could evolve into a more robust and adaptable methodology, helping to bridge the gap between academia and industry. Nevertheless, despite the identified potential, this proposal represents an initial step. Empirical studies are necessary to validate, refine, and expand this vision, and to provide the needed understanding needed to establish more effective pathways for leveraging AI in support of SE research problem formulation.

\section{Conclusion and Future Work}
This paper explored the integration of AI agents into the early stages of Software Engineering (SE) research, using the Lean Research Inception (LRI) framework as a foundation. Through a descriptive scenario in the context of machine learning code maintainability, we illustrated how AI agents can be used with the aim of assisting researchers by pre-filling problem formulation attributes, aligning stakeholder perspectives, refining research questions, simulating multiperspective evaluations, and supporting decision-making.

The descriptive scenario suggests that AI agents can improve the LRI process by reducing cognitive effort, fostering critical reflection, and improving collaboration. Hence, this integration may represent a promising step toward making SE research more context-aware and aligned with real-world challenges. However, this is a vision paper intended to spark early discussion within the community. Naturally, further empirical studies are needed to evaluate its effectiveness and refine recommendations for its practical application.

\bibliographystyle{ACM-Reference-Format}
\bibliography{references}

\appendix
\end{document}